\documentclass[aps,reprint,amsmath,longbibliography]{revtex4-1} %\documentclass[aps,preprint,amsmath,longbibliography]{revtex4-1} 

\usepackage{graphicx}

\usepackage{mathrsfs}
\newcommand{\emf}{\mathscr{E}}

\usepackage{color}					% For editing purposes

\begin{document}

\title{Student and instructor framing in upper-division physics}

\date{\today}

\author{Deepa N. Chari}
\affiliation{STEM Transformation Institute, Florida International University, Miami, Florida}

\author{Hai D. Nguyen}
\affiliation{Department of Natural Sciences, Tennessee Wesleyan University, Athens, Tennessee 37303}

\author{Dean A. Zollman}
\affiliation{Department of Physics, Kansas State University, Manhattan, Kansas 66506}

\author{Eleanor C. Sayre}
\affiliation{Department of Physics, Kansas State University, Manhattan, Kansas 66506}
\email{esayre@ksu.edu}

\begin{abstract}
Upper-division physics students spend much of their time solving problems.  In addition to their basic skills and background, their epistemic framing can form an important part of their ability to learn physics from these problems.  Encouraging students to move toward productive framing may help them solve problems.  Thus, an instructor should understand the specifics of how student have framed a problem and understand how her interaction with the students will impact that framing. In this study we investigate epistemic framing of students in problem solving situations where math is applied to physics.  To analyze the frames and changes in frames, we develop and use a two-axis framework involving conceptual and algorithmic physics and math. We examine student and instructor framing and the interactions of these frames over a range of problems in an upper-division electromagnetic fields course.  Within interactions, students and instructors generally follow each others' leads in framing. %Over the course of the semester students respond to frequent instructor conceptual frames by using more conceptual frames themselves.
\end{abstract}

%\pacs{01.30.lb, 01.40.Fk, 01.40.Ha}

\maketitle

%%%%%%%%%%%%%%%%%%%%%%%%%%%%%%%%%%%%%%%%%%%

%% MAINMATTER

%%%%%%%%%%%%%%%%%%%%%%%%%%%%%%%%%%%%%%%%%%%%

\section{Introduction}

Epistemic framing refers to how an individual perceives a task at hand and determines what knowledge and tools are useful for completing that task \cite{Hammer2004}. This framing is particularly important when students solve problems in a physics course.  By influencing the tools, knowledge, and attitudes of the students, framing has a significant effect on student learning while completing a problem. For example, framing a problem as merely a worksheet to get through engenders different kinds of thinking and problem solving than framing it as an opportunity to make sense of a physical scenario \cite{Scherr2009framing}; framing a problem expansively \cite{Engle2012Expansive} encourages students to seek real-world connections while framing it narrowly encourages them to focus on the mechanics of the problem at hand \cite{Irving2013framing}.  

If, as an instructor, you would like students to connect ideas across representations and still be able to solve end-of-chapter problems mechanically, you are implicitly calling for a balance between conceptual framing and algorithmic framing. If you encourage them to start a problem conceptually, move to algorithmic work, and reflect on it conceptually, you are calling for productive shifting between frames in the course of problem solving.

In this study we focus on students' and instructor's framings in an upper-division physics course in which students work on problem solving in small groups.  This focus is substantially different than a ``difficulties'' approach (e.g. \cite{Pepper1012Observations}) which seeks to characterize the myriad ways students can be wrong about their physics understanding and problem solving.  Difficulties-based approaches to instruction focus on identifying specific conceptual difficulties and designing curricula to ameliorate them; our approach focuses on students' activities and ideas during problem solving and is not tied to specific topics in a course.

Problem solving in upper-division physics courses is often highly mathematical, blending math concepts and physics concepts in complicated, co-generative ways. Part of learning to do physics is learning to shift flexibly between mathematical and physical ways of thinking within the same problem.  In other words, students need to be able to move between physics frames and math frames.  Students' framing when solving math-rich physics problems reveals how they approach the problems epistemically and how they relate physics concepts to mathematical tools and forms \cite{Bing2009warrants}.   Thus, students' framing influences their problem solving strategies, whether they enjoy instructional activities, and what they learn \cite{Hutchison2009,Engle2012Expansive,Gupta2013a}.  Because instructors' framing can affect students' framing \cite{Irving2013framing,Scherr2009framing}, we are also interested in the instructor's effects on student framing, both in the moment of the interaction and over the course of the semester. 

We are interested in how students solve problems in natural settings as part of their normal development as physicists. This interest steers us to observe groups of students solving problems during class. In this study, we investigate how students' epistemic framing during problem solving affects and is affected by the instructor's epistemic framing.

We build on prior work identifying frames for classroom observations and math-in-physics problem solving at the upper division. Other studies have developed a theoretical framework and coding for classroom observations in electromagnetic theory \cite{Hutchison2009,Engle2012Expansive,Gupta2013a,Nguyen2016StuFraming} and have shown that it provides an underlying structure to student difficulties in quantum mechanics \cite{Modir2017QuantumDifficulties}.  In this study, we have two goals. The first is to find how prevalent each frame is for both students and an instructor in electromagnetic theory, how much time they typically spend in each of them, and what frames they are in immediately before and after any given frame. Secondly, we investigate how the instructor's and students' frames interact and influence the frames of each other.

In the following sections, we detail the instructional context of our work, review prior work on epistemic frames, elaborate the math-in-physics frames that we use, and present data on student and instructor framing. We conclude with implications for future research and instruction.

\section{Context of study}

We collected data from two classes (Fall 2013 and Fall 2015) of an electromagnetic theory (E\&M) course taught by the same instructor.  This upper-division course covers chapters two through seven of Griffiths' textbook \cite{Griffiths1999EM}.  Each class enrolled fifteen to twenty students.

The course meets for four fifty-minute sessions each week. Class sessions are highly interactive and involve a significant amount of time during which students solve problems in groups of three or four. Students are free to choose their groupmates and often stay in the same group throughout the course.

Each session follows one of three scenarios: multiple problem solving, extended problem solving, or tutorial.  In a multiple problem solving session, students spend the majority of their time working on two to three problems on a single topic.  In an extended problem solving session, students work on one longer problem for the majority of the period. In a tutorial session, students solve tutorial problems \cite{Baily2013a} in groups.  In all scenarios, the instructor does very little lecturing.  Problem solving time dominates the class.

We divide the students' problem solving time into two subcategories: \textit{Collaborative problem solving} where students discuss together, solving problems on their shared whiteboard; and \textit{Individual problem solving} where students work independently on problems without interacting with each other much.  In both cases, the instructor briefly visits the groups to check on progress and provide necessary assistance. A typical class also includes short lectures by the instructor (Lecture time) and administrative activities such as homework collection or announcements (Admin time). 

\begin{figure}[thp]

\begin{center}

\includegraphics[width=1.0\linewidth]{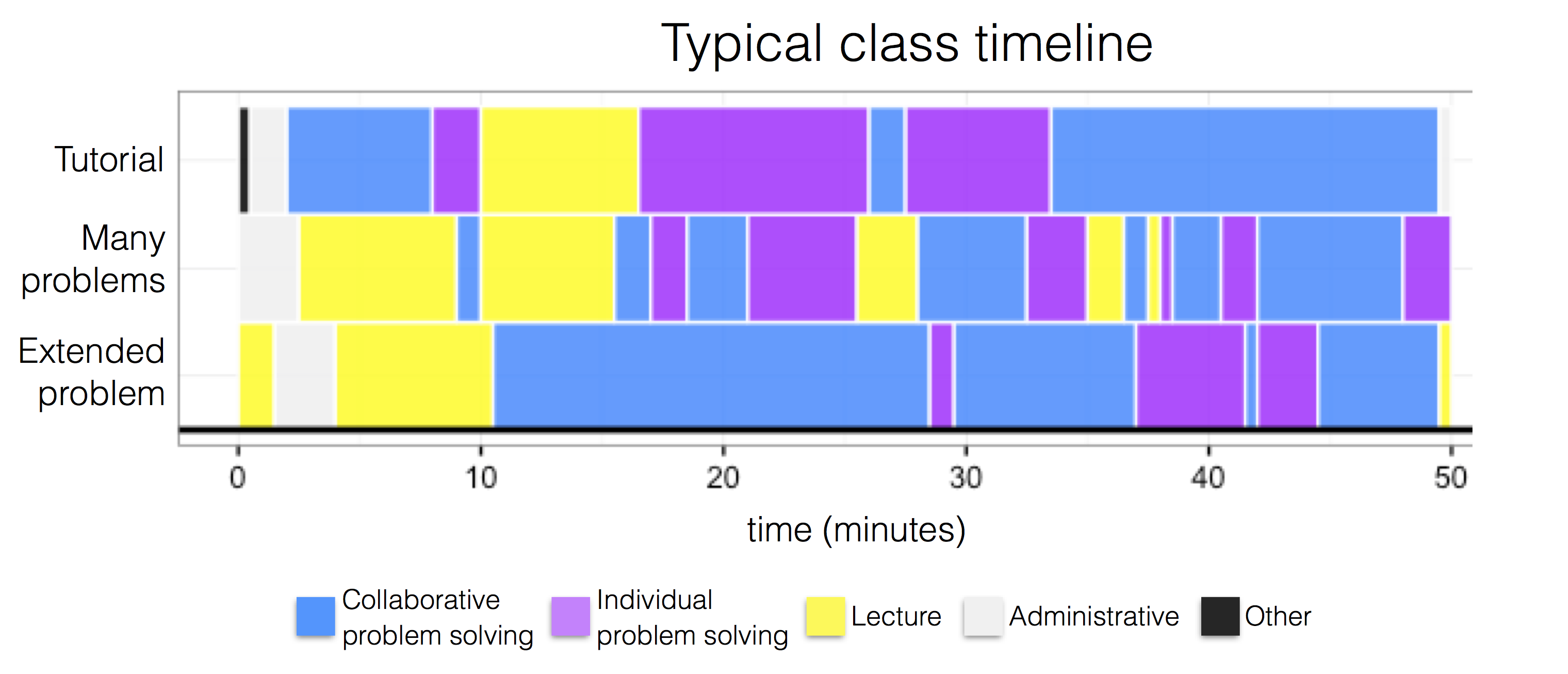}

\end{center}

\caption{Distribution of time in a typical class session of each scenario.  On extended problem solving days, students work on one longer problem for most of the class.  On multiple problem solving days, students work on several problems in series.  On tutorial days, students work through a guided series of problems on a worksheet. Each of these distributions was measured in a single representative class day of its respective type.}

\label{fig:typicaldays}

\end{figure}

Figure \ref{fig:typicaldays} shows how time spent on each kind of activity is distributed in each scenario (as calculated from a single class meeting; this distribution is typical of these kinds of class meetings). We can see that, for all three scenarios, students spend most of the time solving problems, either individually or in collaboration within the group. On both extended problem solving days and multiple problem solving days, students work interactively in their groups extensively.  On tutorial days, they also spend substantial blocks of time working on the tutorial but not interacting with each other very much. 

The instructor's lectures occur mostly at the beginning of the session or a short time into the session after students have had an initial look at the problems. In a multiple problem solving session, lecture occurs several times throughout the session, since it is needed to direct students and provide information to work on each new problem.

\section{Theoretical framework}

Ample work in physics education research has used the epistemic framing perspective  \cite{Bing2009warrants,Wolf2014JustMath,Hammer2004rft}. These studies include several investigations on problem solving in small groups \cite{Scherr2009framing,Tuminaro2007,Irving2013framing}.
%A few framing schemes have been used for upper-division physics content, particularly math-in-physics, and for classroom observations.  

Scherr and Hammer investigated student behavior during introductory level tutorial activities and defined four frames: worksheet, joking, TA, and discussion\cite{Scherr2009framing}. This model separates discussion from other types of communication (such as joking and interacting with the instructor), and excludes the TA from participating in discussion. However, in an upper-division classroom, the instructor often collaborates with students to frame a problem. Therefore, students' discourse while interacting with the TA or the instructor may also be productive. In this study, we are interested in how instructor framing affects student framing (and vice-versa), so we need a finer level of detail than this scheme provides.
%Since the focus of our study is to explore the finer details of math and physics discussions, we consider all discussion about the problem, whether serious or humorous, as relevant to the framing\cite{Irving2013,Wolf2014}.

An alternative approach to discrete frames is to define a coordinate plane of framing and allow students' discussions to roam freely in the plane.  \citet{Irving2013framing} break students' frames along two axes: expansive to narrow and serious to silly.  The expansive to narrow axis focuses on the scope of the discussion while the serious to silly axis focuses on how students engage in the discussion. In another study, \citet{Wolf2014JustMath} overlap this model with Scherr and Hammer's ``worksheet'' frame to define a ``just-math'' frame. In the ``just math'' frame, students' discourse is restricted to check-ins and sharing their expectations which is not typical of the worksheet frame. Student discussions are brief and happen before they proceed to do the final mathematics of the problem. Just-math is a useful epistemic frame; however, it does not tell us if the check-ins are about physics ideas or mathematical constructs. Such details are necessary to understand the interconnected roles of conceptual physics ideas and mathematical tools.

Bing and Redish proposed four epistemic frames for mathematics used in physics contexts\cite{Bing2009warrants}. They are: authority, calculation, mathematical coherency, and physical mapping.  The authority frame describes students using notes, prior results, or instructor arguments without any need for verification. However, in a classroom the instructor's interaction with students is not simply to provide facts and information. Nudging and scaffolding also contribute in meaningful ways to problem solving. The authority frame does not consider these factors and is ill-suited for analyzing classroom discourse.

In the mathematical coherency frame, students tend to look for mathematical coherency between two superficially different equations. In the physical mapping frame, students describe the physical or geometric features of a system and to coordinate them with suitable symbolic or algebraic representations. The calculation frame deals with students focusing on correctly following mathematical procedures. In this frame, students simplify mathematical equations, or perform ``plug and chug'' types of activities to attain solutions. However, students may employ conceptual knowledge of mathematical constructs to obtain a result or apply known methods to approach the next level of computation\cite{Sfard1991}. The calculation frame does not distinguish between algebraic computation and conceptual manipulation of mathematical constructs, yet this difference is important to problem solving at the upper division.

In our prior work in this course\cite{Nguyen2016StuFraming} and an interactively-taught quantum mechanics course\cite{Modir2017QuantumFraming}, we developed a framework for investigating students' framing along two axes:  math to physics and conceptual to algorithmic.  This math-in-physics framework has several affordances for our context and research questions. Unlike the Scherr and Hammer scheme or Bing and Redish scheme, it can be applied equally to students and instructors (as can the Irving \textit{et al.} scheme). However, the Irving \textit{et al.} scheme does not focus on the mathematical or physics content of students' discussion, and thus does not match our research questions well.   In contrast, the Bing and Redish scheme was developed for math-in-physics issues but did not distinguish between conceptual math and algorithmic math nor did it allow for analysis of instructor framing as well. Our proposed scheme is a better match to our research questions and observational data than prior schemes.

\begin{figure}[tbh]
\begin{center}
\includegraphics[width=1.0\linewidth]{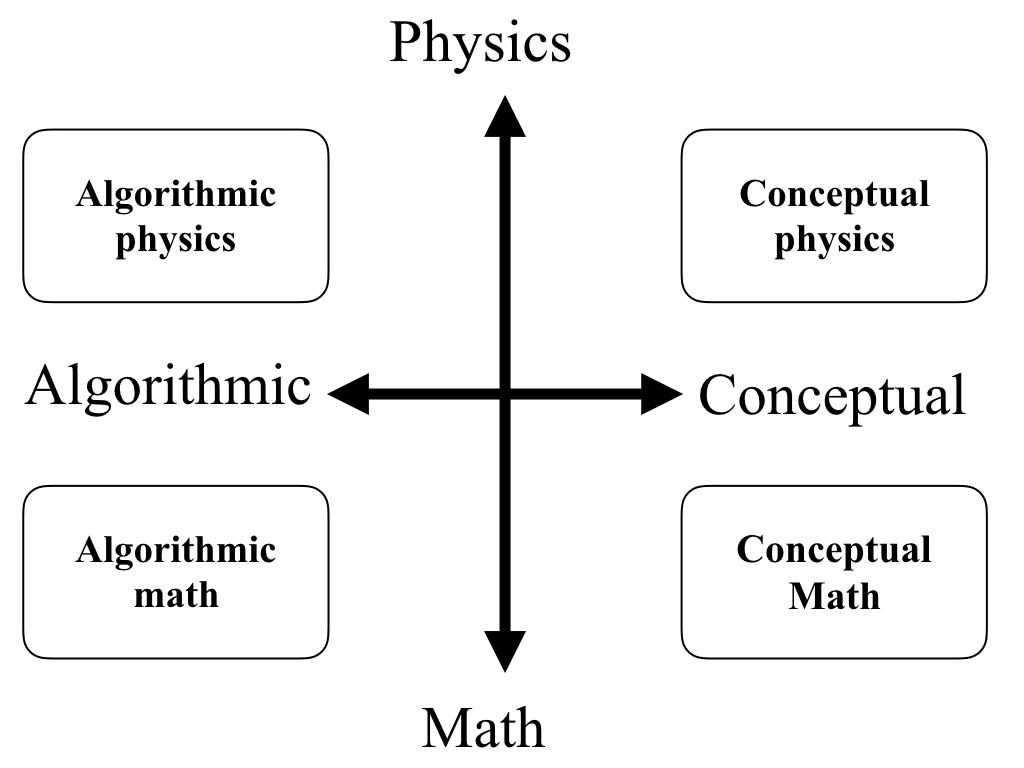}
\end{center}
\caption{Schematic of math-in-physics framework showing two axes and four frames.}
\label{fig:framework}
\end{figure}

%\subsection{The four framing frames}

The two axes in this framework create four quadrants (Figure \ref{fig:framework}) corresponding to four frames that students and instructor may take on during problem solving. The characteristics of each frame are discussed below.

\begin{description}
\item[Conceptual physics (CP) frame]
Students and instructor are in this frame when they discuss physics scenarios or phenomenon, about properties of physics quantities related to the task at hand. They may also exploit the symmetry of a physical system by investigating related concepts.
\item[Algorithmic physics (AP) frame]
In the algorithmic physics frame, students recall physics equations or apply physics knowledge to rearrange known equations using math. Students may also derive expressions for specific cases from a general physics equation or validate an expression via dimensional analysis.
\item[Algorithmic math (AM) frame]
This frame refers to performing mathematical computation by following well-established protocols without questioning the validity of those protocols, e.g. solving an equation or computing an integral.
\item[Conceptual math (CM) frame]
Students are in this frame when they exploit properties of mathematical constructs to quickly obtain a result without diving into algorithmic manipulation, e.g. noticing that all the odd terms in a sum are equal to zero.
\end{description}

Elsewhere,\cite{Nguyen2016StuFraming,Modir2017QuantumFraming} we've exhaustively detailed these four frames and linked them to student discourse.  To illustrate them here, we offer a brief example from electrostatics. We chose this example to illustrate how different solution paths use different frames.  While all three paths yield the same solution, the reasoning and framing for each path is substantially different. 

Suppose you have a spherical shell of radius $r$ which carries surface charge density $\sigma = \alpha \sin(2\theta)$, where $\alpha$ is a constant and $\theta$ is the polar angle (from the positive z-axis).  What is the total charge on the sphere?

\begin{figure}[htb]
\begin{center}
\includegraphics[width=1.0\linewidth]{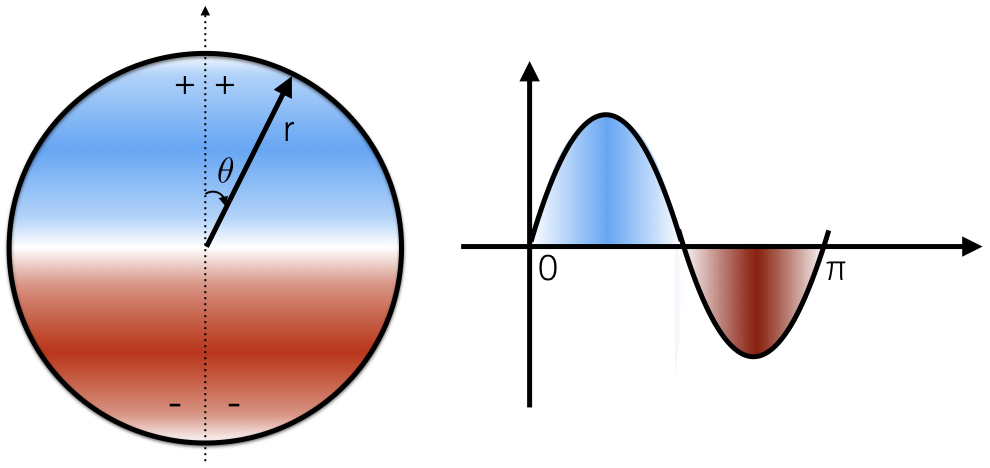}
\end{center}
\caption{The \textit{charge map} solution to the charge sphere problem, illustrating the conceptual physics frame. On the left, a spherical shell with surface charge density $\sigma = \alpha \sin(2\theta)$; on the right, the function $\sin(2\theta)$ from $0$ to $\pi$.}
\label{fig:chargemap}
\end{figure}

Using the conceptual physics frame, we can map the sine function to a globe where the northern hemisphere holds positive charge and the southern hemisphere holds an equal amount of negative charge.  The \textit{charge map} solution (figure \ref{fig:chargemap}) says that the charge on the two hemispheres adds to zero.    

Alternately, one could take a mathematical approach, noting that the total charge is the integral of the charge density over the surface, $Q = \int \sigma dA$, where the area element $dA = {r}^2 \sin(\theta) d\theta d\phi$.  This yields the integral equation
\begin{align}
Q =  \int \alpha \sin(2\theta) {r}^2 \sin(\theta) d\theta d\phi
\end{align}
Two solution paths present themselves.  In the algorithmic math frame, one could use a long series of trigonometric identities to simplify the two sine functions and solve the integral.  In the conceptual math frame, one could exploit the orthogonality of sine functions and skip directly to the solution: the integral equals zero. In either case, the path is mathematical and does not rest on physics ideas like charge. 

Most problems in upper-division physics are longer than this brief example.    If this problem were be the start of a much longer problem to find the electric potential far away from the sphere, students might approach the longer problem in the algorithmic physics (AP) frame.  In the AP frame,  they might apply an multipole expansion algorithm: checking for appropriate symmetry, finding the potential on axis, building a series solution, matching boundary conditions, etc.  %Someone who treated using the multipole expansion as an algorithm to apply would be in an algorithmic physics (AP) frame.  
 %The fourth frame in our framework, algorithmic physics, does not appear in these solution paths.  It might occur if students were to follow a prescriptive series of problem solving steps by rote.  
To distinguish prescriptive or rote problem solving from sense making or conceptual thinking, we need to examine students' discourse including language, tone, and prosody; for this reason, we prefer to use video-based data in determining students' frames. 

In longer or more difficult problems, students are likely to transition between frames as different phases of problem solving require different kinds of thinking in order to be productive. Unlike some prescriptive problem solving rubrics (e.g. \cite{Heller1992a,Wilcox2015Dirac}), we don't make normative recommendations about an \textit{a priori} sequence of frames to transition among; the details of which frames are productive must be an interaction between the students and the problem at hand. %; however, we note that when students get stuck in problem solving, 

\section{Methodology}

%\subsection{Selection of method}

In this study, we seek to understand students' and instructor's framing in real classroom settings. Research studies on epistemological framing often use video as the primary source of data due to its ability to capture rich information about students' behavior and discourse\cite{Bing2009warrants,Irving2013framing,Scherr2009framing}.   Following the coding scheme developed in prior work\cite{Nguyen2016StuFraming}, we analyzed 50 episodes of problem solving drawn from 30 weeks (about 110 instructional days, excluding exam days) and 440 hours of video, in which 3-4 groups were videoed for the whole of every period (save for occasional technology hiccups). In addition to the codes for the four frames, we also included a series of codes for ``other'' behavior, which encompasses administrative work like handing in homework or off-topic discourse like socializing.  This behavior is discussed in section \ref{sec:ocodes}.

In selecting episodes for analysis, we selected for both a wide range of topics (spanning the whole 15-week course as in Table \ref{tab:distribution}) and multiple groups per topic to increase the breadth and depth of our study.  We also looked for high quality interactions: ones in which students made substantial progress in problem solving, and where their work is both visible and audible. The episodes range from three to 25 minutes in length, and are exclusively drawn from collaborative problem solving and tutorial activities during class.

\begin{table*}[th]
\caption{Distribution of 50 episodes across the course topics}
\centering
\begin{tabular}{|c|c|c|c|c|c|c|c|}
\hline
Topic &E in vacuum &E in matter &E and B &B in vacuum &B in matter &Waves &Co-axial cables / circuits\\
\hline
No. of examples &14 &5 &5 &7 &3 &1 &1\\
\hline
No. of groups &12 &3 &9 &12 &3 &2 &1\\
\hline
No. of examples per topic &12 &11 &9 &12 &3 &2 &1\\
\hline
\end{tabular}
\label{tab:distribution}
\end{table*}

%\subsection{Inter-rater reliability}

For selected episodes, two researchers independently assigned codes and determined the time spent in each frame. The raters were provided with detailed instruction about the code book and coding scheme. Three episodes were selected for training purposes and were not included in the data. The raters compared beginning and end times and the code assigned to that period of time. A difference of $\pm5$ seconds for the beginning or end times was treated as a separate time code.

During the training session, inter-rater reliability on framing codes, beginning and ending times was 95\%, 84\% and 84\% respectively. Discrepancies were discussed in a follow-up meeting before starting the reliability test in which 14 episodes randomly selected from our pool of 50 episodes were tested. After training and discussion, the inter-rater reliability for framing codes, beginning and ending times were 95\%, 93\% and 93\% respectively.

\section{Framing in one episode\label{sec:example}}
\begin{figure}[tb]
%\begin{center}
\fbox{
\begin{minipage}{.9\linewidth}
The infinitely long outer solenoid (radius $b$, turns per length $n_o$) carries time-varying current $I_{outer} = I_0 \sin(\omega t)$.  Find the EMF in the inner, finite solenoid (radius $a$, total turns $N_i$).

\includegraphics[width=.8\linewidth]{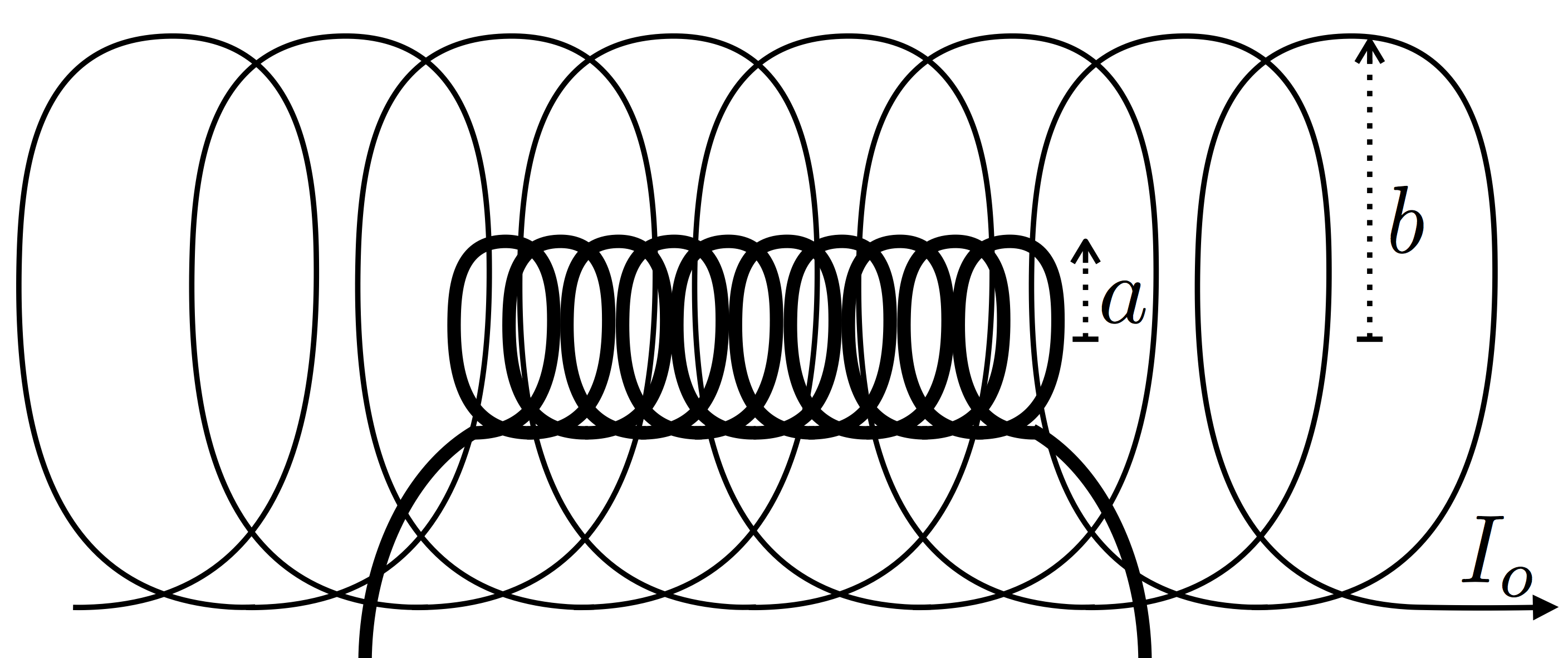}

Solution:

Magnetic field $\mathbf{B}$ in outer solenoid:
\begin{align*}
\mathbf{B}_{outer} &= \mu_0 n_{o} I_{outer}\\
&=\mu_0 n_{o} I_0 \sin(\omega t)
\end{align*}

Magnetic flux $\Phi$ through inner solenoid:
\begin{align*}
\Phi &=  N_i \int \mathbf{B}_{outer}\cdot d\mathbf{A}_{inner}\\
&= N_i \int_0^{2\pi} \int_0^a \left( \mu_0 n_{o} I_0\sin(\omega t) \right) \left( r dr d\theta \right)\\
%&= \left( \mu_0 n_{o}\sin(\omega t) \right) (2\pi) \int_0^a rdr
&= N_i \pi a^2 \mu_0 n_{o} I_0\sin(\omega t)
\end{align*}

EMF $\emf$ in inner solenoid:
\begin{align*}
\emf & = - \frac{d\Phi}{dt}\\
&= -(N_i \pi a^2 \mu_0 n_{o} I_0)\frac{d}{dt}(\sin(\omega t))\\
&= -(N_i \pi a^2 \mu_0 n_{o} I_0)(\omega \cos(\omega t))
\end{align*}

\end{minipage}
}
%\end{center}
\caption{Solution to the nested solenoids problem.}
\label{fig:solenoid}
\end{figure}

In this section, we examine the frames and frame shifts for one group of students and the instructor in one problem solving episode.  Matt, Abbey, Jessy, and Chen\footnote{Student names are pseudonyms.} are solving a problem about two nested solenoids. Students have to find the induced electromotive force in the short inner solenoid, given the changing current $I_{outer}$ in the outer infinite solenoid (figure \ref{fig:solenoid}).

%\subsection{Episode description and narrative}

\begin{figure}[tbh]
\begin{center}
\includegraphics[width=1.0\linewidth]{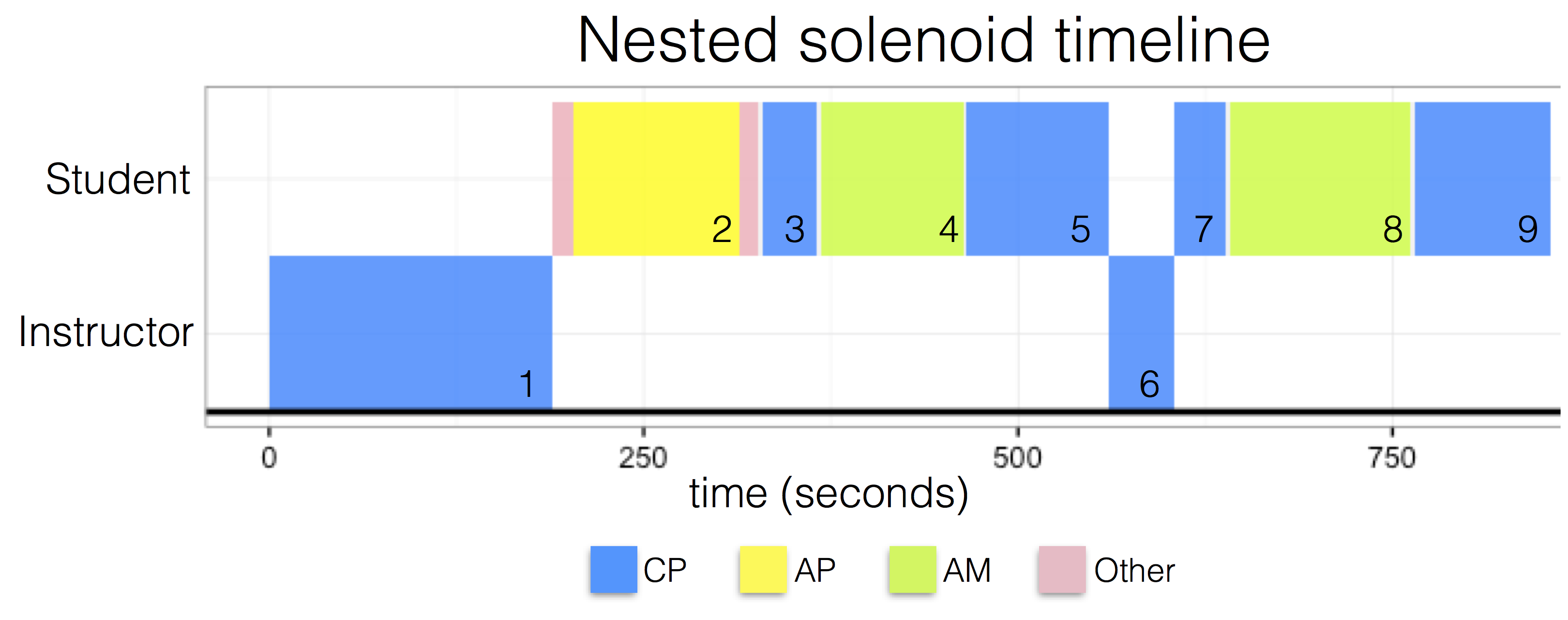}
\end{center}
\caption{Timeline (in seconds) of the nested solenoids episode.  Upper row: student framing.  Lower row: instructor framing.  Segments of the timeline are marked (\#1-9), corresponding to segments in the text.}
\label{fig:timeline}
\end{figure}

The timeline plot  (figure \ref{fig:timeline}) tracks the students' and the instructor's framing as they progress through the problem solving process. We present a brief narrative and some excerpts of transcript to discuss students' framing in the episode.

\textit{Segment \#1} (seconds 0-189), After posing the problem, the instructor gives a brief lecture about induced EMF in a solenoid. She suggests that students make a plan for solving the problem. This requires students to think about the physical system and phenomenon involved, so the instructor is in the conceptual physics (CP) frame. Before starting the problem, the students are quiet for 14 seconds (seconds 189-203)

\textit{Segment \#2} (seconds 203-314), Matt recalls the formulas for magnetic field and flux , arguing that ``the magnetic field is $\mu_0$ times the current \dots So, remember the flux is $B$ field integrated over the area.'' As he speaks, he flips through his textbook. Matt's discourse indicates that he was recalling relevant equations and the group follows him. Up to this point the group has not explicitly charted a plan to solve the problem as suggested by the instructor. Therefore we interpret that the group was in algorithmic physics (AP) frame.

\textit{Segment \#3} (seconds 327-369), After a short gap of silence (seconds 314-327), Matt shares his conceptual understanding of geometry and flux in short bursts with long pauses between them.
\begin{description}
\item[Matt] \dots The flux is gonna be the same in the small solenoid \dots from there you can calculate the EMF  \dots because, you are inducing an EMF in the smaller [::] solenoid, by producing flux in the big [::] solenoid.
%\item[Matt] 
\dots So that's, um, really, we are looking at the flux in the little solenoid, right?
\end{description}
Until this point, Matt is the only speaker in the group. The other students in the group agree with him by nodding.  We interpret their nodding as agreement with both his argument and his framing. We interpret this section as conceptual physics (CP).

%Between marks \#2 and \#3, the group shifts from an algorithmic physics frame to a conceptual physics frame.  

\textit{Segment \#4} (seconds 369-463), Abbey and Chen start working on Matt's proposal and begin to identify the area to be used in the flux formula.  They check in with each other about the values of the quantities that they're calculating, and mostly they are focused on writing and solving equations.
\begin{description}
\item[Abbey] So, it's $B$ dot $dA$, right (pauses), but which $dA$?
\item[Matt] The smaller one. Shall we be writing this on the board?
\end{description}
Matt starts writing the formulas on the whiteboard. Abbey asks a few questions about the radius of a solenoid. She also prompts Matt with information on how to solve the equations. They compute the area integral and set up a derivative. The group is in the algorithmic math (AM) frame because they are trying to perform algebraic procedures without talking about the conceptual meaning behind them. 

\textit{Segment \#5} (seconds 465-561), Jessy stops working and asks if the region of interest should be the outer loop. Matt explains to her the idea of magnetic flux and the region of interest. To support her question, Jessy draws everybody's attention to the details of the problem, quoting the problem statement: 
\begin{description}
\item[Jessy] [reading] ``Find the flux of $B$ for the big [::] loop'' \dots that's one of the instructions.
\end{description}
The discussion that follows from Jessy's statement causes a shift away from algorithmic computation. The group goes back to discussing physics concepts to determine the area of interest. During the group's conceptual discussion, the instructor stops by and starts observing this group.  

\textit{Segment \#6} (seconds 561-605), The instructor helps the students determine the region of interest  by explaining how the outer solenoid creates flux in the inner solenoid. 
\begin{description}
\item[Instructor] Emmh, yeah that is somewhat ambiguous. So, we need the flux created by big loop in the region of the small loop. Because that's the only way we know how to use Faraday's law. The flux in the small loop will tell the EMF around the solenoid. But, that flux is created by big loop.
\item[Abbey] So [pause] we will find the flux on the inner loop.
\end{description}
We interpret this discourse as being in the conceptual physics (CP) frame.

\textit{Segment \#7} (seconds 606-642) At the start of this segment, the instructor leaves the table.  The students continue in the conceptual physics (CP) frame after she leaves, discussing the concept of flux in a ring caused by the magnetic field of a solenoid. 
%    \begin{description}
%    \item[Matt] Ok, so you have to take a look at what we are ultimately solving it for. The EMF in the small loop. And the EMF in the small loop is only [::] generated by the flux in the small loop. And where [::] does that flux come from is the flux from the big loop.
%    \item[Abbey] So, is it correct to say that emmh we take magnetic field from big loop dot $dA$ for the small loop.
%    \item[Matt] Yeah, as that's what we are concerned about.
%    \end{description}

\textit{Segment \#8} (seconds 642-762) Once students confirm that the flux is produced due to the outer solenoid, they shift to algorithmic math (AM) to obtain a final expression.

\textit{Segment \#9} (seconds 765-855) At the end of this episode, the group arrives at an answer and continues the discussion about the directionality of the EMF. They compare finite and infinite solenoids and add the term `N' to the solution to indicate the finite number of turns for the given solenoid. Both of these discussions are also in conceptual physics (CP) frame.

%\subsection{Time analysis and discussions}

\begin{figure}[tbh]
\begin{center}
\includegraphics[width=\linewidth]{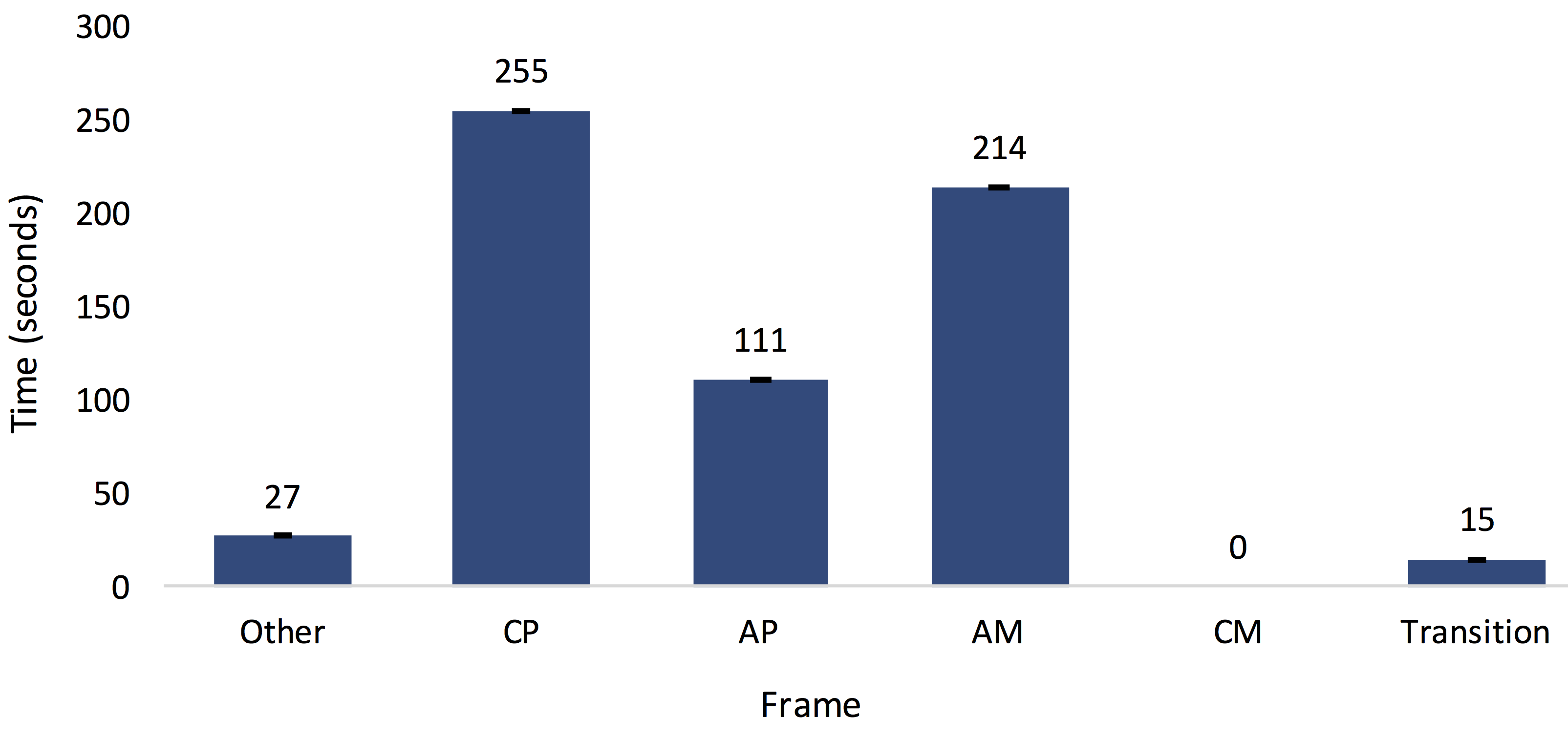}
\end{center}
\caption{Time (in seconds) that the students spend in each frame in the nested solenoid example.}
\label{fig:time-solenoid}
\end{figure}

This problem requires students to examine the physical system to understand the physics phenomenon involved, to recall relevant formulas, and to perform some algorithmic calculations. Students spend time in the  conceptual physics (CP), algorithmic physics (AP), and algorithmic math (AM) frames (Figure \ref{fig:time-solenoid}).  Though they return to conceptual physics four times (marks 3, 5, 7, 9), they only spend an average of 64 seconds each time there.  In contrast, they enter the algorithmic physics (AP) frame once, for 111 seconds, and the algorithmic math (AM) frame twice, for an average of 107 seconds each time.   The instructor spends all her time in conceptual physics (CP).

\section{Multiple episodes}

We expand our scope to all 50 episodes to examine the instructor's and students' framing quantitatively. We determine the total time spent in each frame, the number of episodes of each frame, and average time per episode in each frame for 50 episodes as a whole. The total time that the students spent in all frames was just under 6 hours (5h42m). The instructor spent a total just over 2 hours (2h05m) in all frames. Note that the breakdown of framing in episodes is not reflective of the total class time because we've selected episodes for periods of student problem solving, not extended whole-class lecture or administration.

\begin{figure}[tb]
\begin{center}
\includegraphics[width=0.45\linewidth]{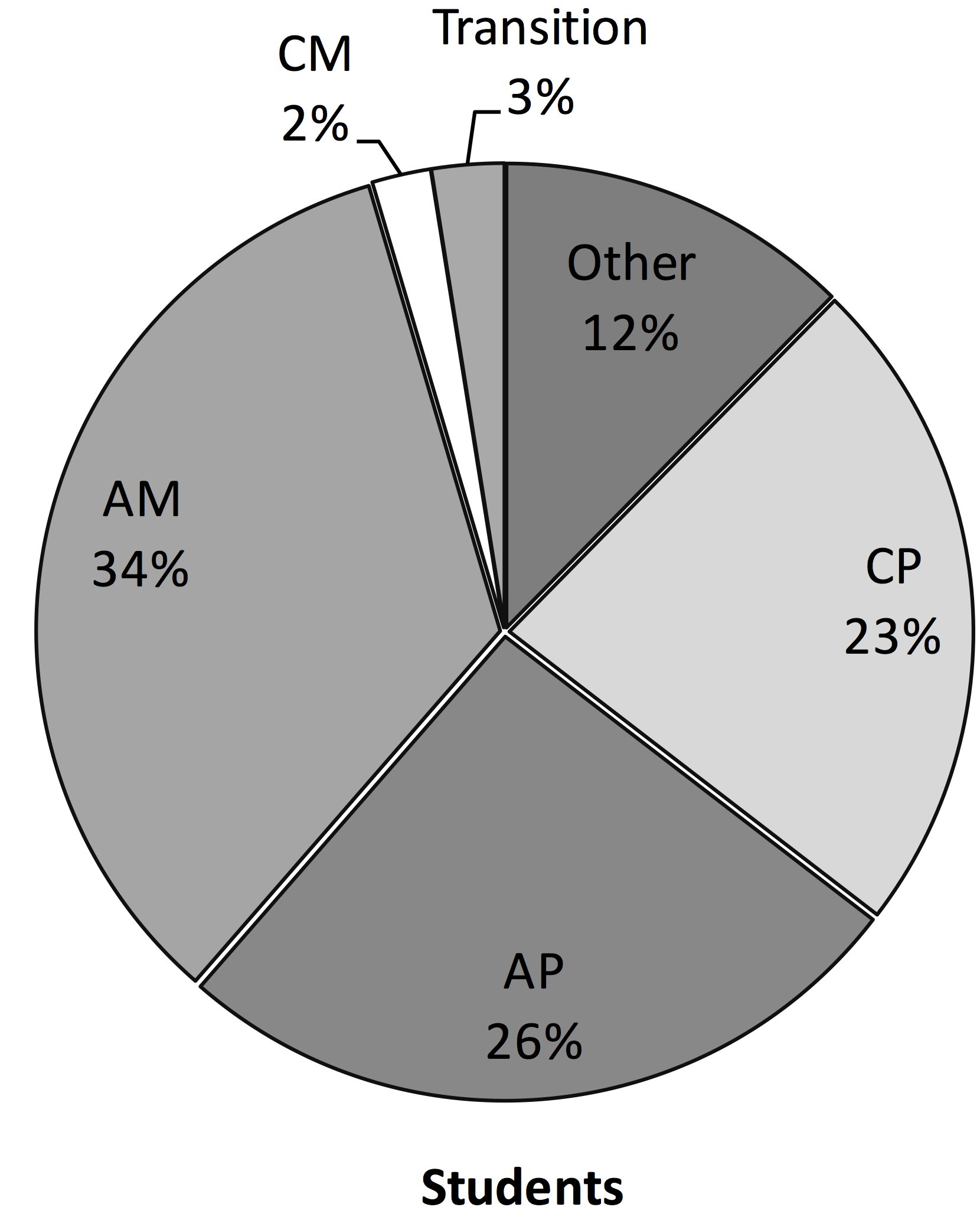}
\includegraphics[width=0.45\linewidth]{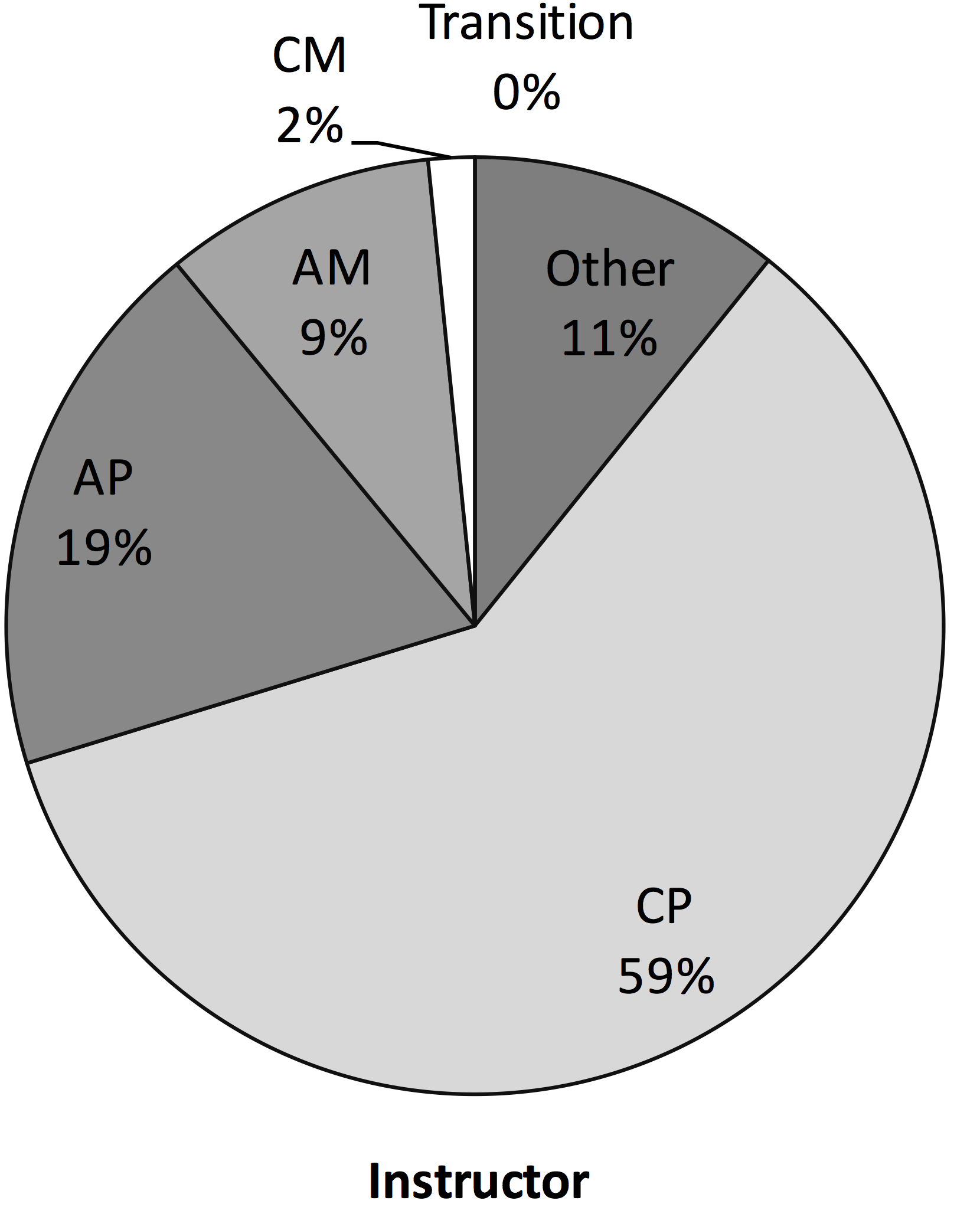}
\end{center}
\caption{Total time (in \%) in each frame across 50 episodes.  Left: students; right: instructor}
\label{fig:pies}
\end{figure}

Overall, students spend the greatest fraction of their time (34\%) in the algorithmic math (AM) frame; in contrast, the instructor spends most of her time in the conceptual physics (CP) frame (figure \ref{fig:pies}) when interacting directly with student groups during problem solving. Both students and instructors spend the smallest fraction of their time in the conceptual math (CM) frame (~2\%), meaning that it is extant, but rare.  

Across 50 episodes (Table \ref{tab:timing}), students enter the algorithmic math (AM) frame 78 times for an average of 90 seconds each, but the conceptual math (CM) frame only 6 times, averaging 70 seconds each.   The instructor is less inclined than the students to algorithmic math (AM), entering it 24 times (average 29 seconds each time), but not substantially more disposed than the students to conceptual math, entering it 4 times (average 30 seconds each time).  

\begin{table}[tbhp]
\caption{Across all 50 episodes, the time that students (left) and instructors (right) spend in each frame.  For each frame, total time (h:mm:ss), counts, and average time per count (t/c) (s) are presented.}
\label{tab:timing}
\begin{center}
\begin{tabular}{|l|ccc|ccc|}
\hline
 & \multicolumn{3}{|c|}{Student}  & \multicolumn{3}{|c|}{Instructor} \\
Frame & time & count & t/c & time & count & t/c\\
\hline
Other & 0:42:18 & 73 & 35s & 0:13:28 & 10 & 81s\\
CP & 1:18:59 & 95 & 50s & 1:14:20 & 58 & 77s\\
AP & 1:28:52 & 90 & 59s & 0:23:27 & 20 & 70s\\
AM & 1:56:26 & 78 & 90s & 0:11:43 & 24 & 29s\\
CM & 0:07:01 & 6 & 70s & 0:02:00 & 4 & 30s\\
Transition & 0:08:38 & 75 & 7s & 0 & 0 & -\\
\hline
\end{tabular}
\end{center}
\end{table}%

This suggests that students are more inclined to performing algorithmic computations than thinking conceptually about mathematical ideas. Alternately, the problems selected for our study -- while representative of the problems in upper-division E\&M -- may not include much need for the conceptual math frame.
%a lot of mathematical constructs on which students can take a shortcut to the result. 

This paucity of conceptual math is not meant as an indictment of the pedagogy or problems in the course.  Instead, we count it as a success for theory-driven research: it's common in the literature to ascribe predominately computational or algorithmic ideas to math frames(e.g. \cite{Bing2012journeyman,Redish2006math}) reserving conceptual thinking for physics alone, but our theory predicts the existence of a conceptual math frame distinct from an algorithmic one.  

Results from a one-way ANOVA with Bonferroni \textit{post hoc} test show that the average time per count in the algorithmic math (AM) frame is significantly greater than that in all other frames ($p < 0.05$) for students. This could be due to the length of the math procedure, the challenges of using complex tools, and/or that students have to write their calculation on the whiteboard. In our episodes, we see that students spend the majority of their time doing math to find the answer. Discourse happens in short bursts since students tend to work independently and occasionally check in with the group to validate their work.

The instructor spends comparatively little time for each time she enters the math frames. We interpret this difference as the instructor's unwillingness to perform tedious mathematical computations, coupled to her strong preference for connecting students' work to conceptual physics ideas.  Her most common framing is conceptual physics (CP), entering it 58 times for 77 seconds each (on average), and covering 59\% of her framing.  This is roughly 3 times more common than her next-most-common frame, though of comparable average duration to algorithmic physics (AP) frame (average 70 seconds each time).  

\section{Frame shifts}

As students solve problems, they shift from one frame to another, moving about in the plane.  To better understand the nature of how the instructor's framing affects student framing (and vice versa), we investigated the distribution of preceding frames as a function of the distribution of following frames.  

\subsection{Shifts involving two frames}
\begin{figure}[tb]
\begin{center}
\includegraphics[width=.9\linewidth]{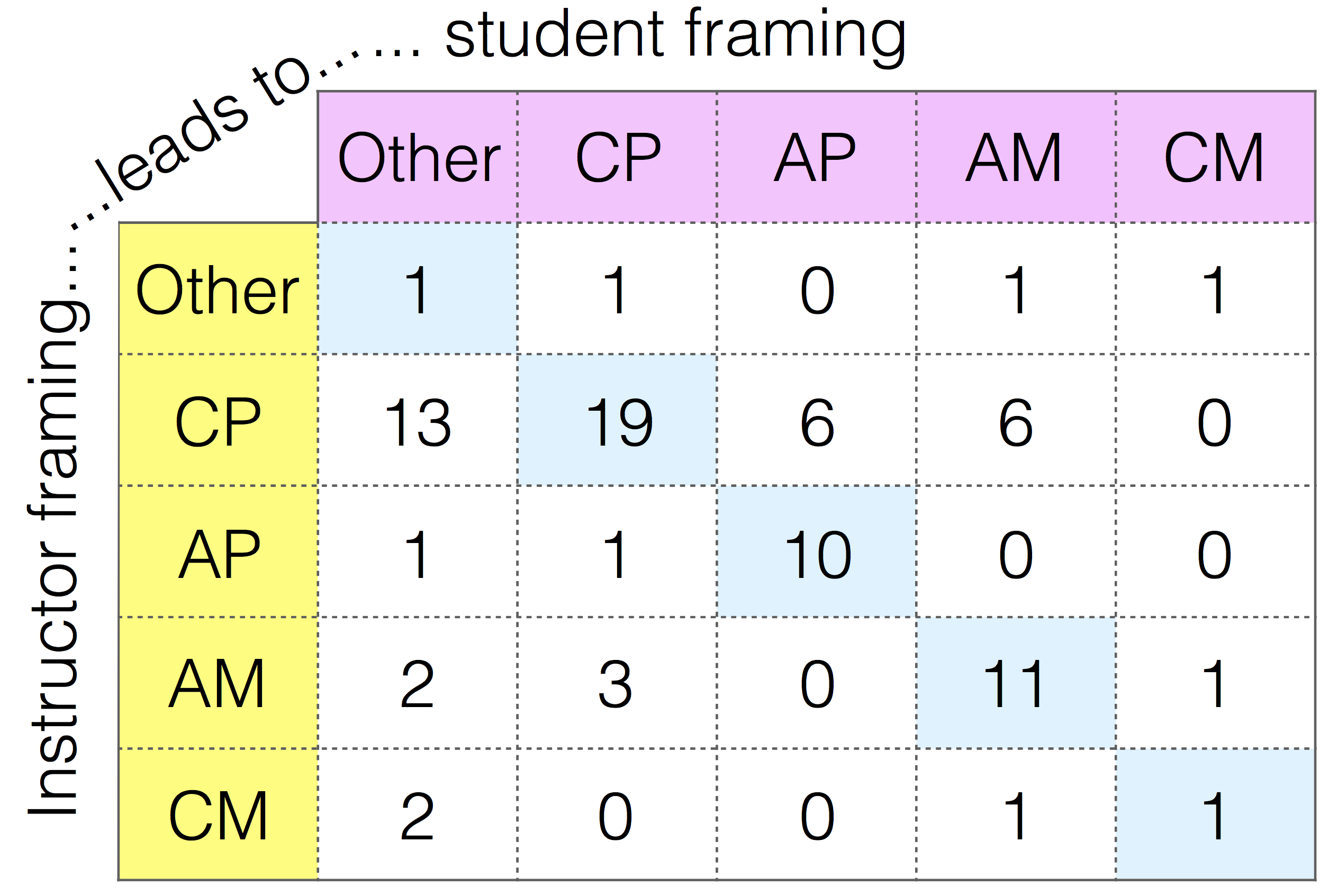}
\end{center}
\caption{Instructor-Students shift matrix.  The instructor's framing (rows, yellow) leads to the students' framing (columns, purple).  Diagonal elements are highlighted in blue.}
\label{fig:IS}
\end{figure}

\begin{figure}[tb]
\begin{center}
\includegraphics[width=.9\linewidth]{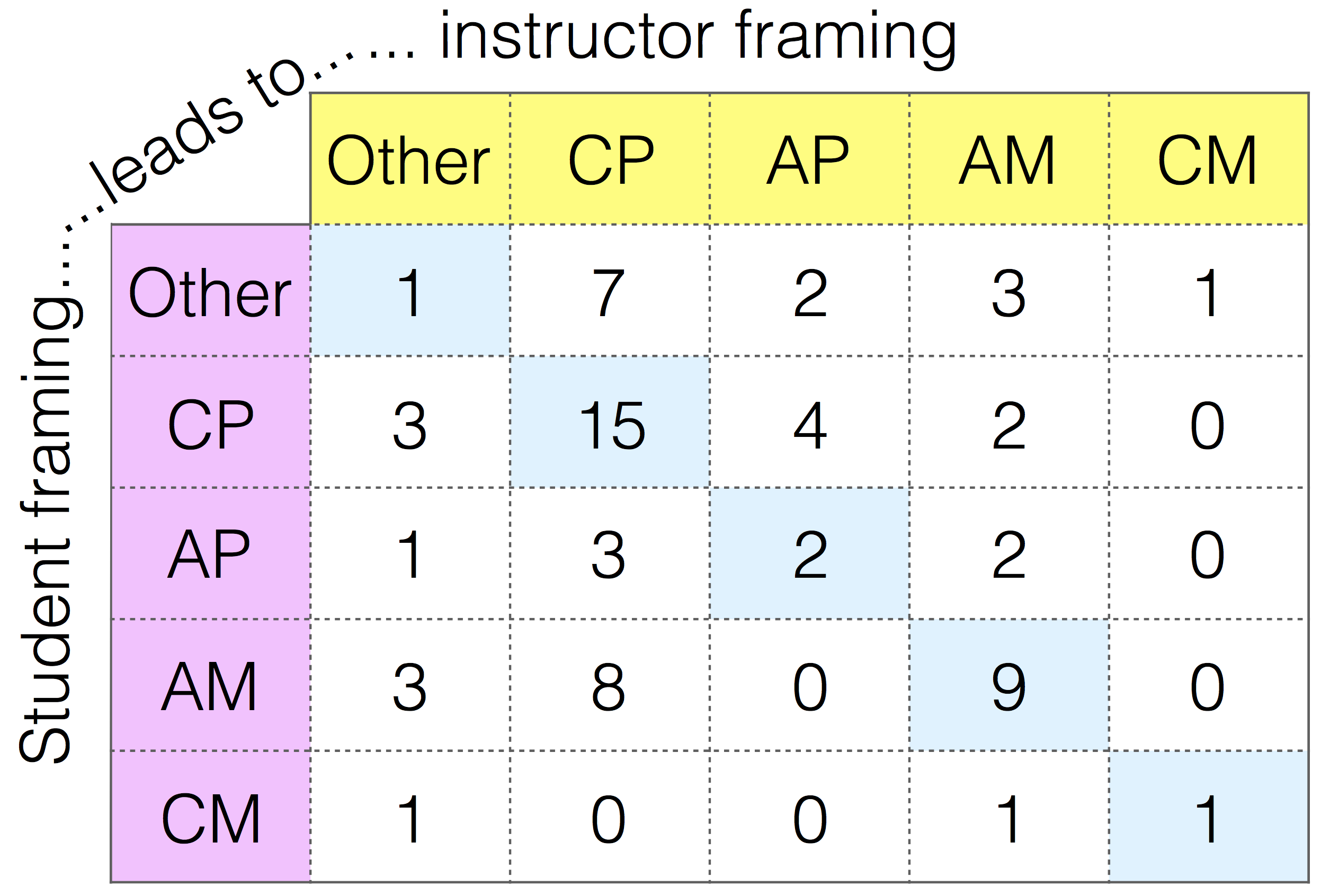}
\end{center}
\caption{Students-Instructor shift matrix. The students' framing (rows, purple) leads to the instructor's framing (columns, yellow).  Diagonal elements are highlighted in blue.}
\label{fig:SI}
\end{figure}

To visualize how preceding frames affect following ones, we generate transition matrices (figures \ref{fig:IS} and \ref{fig:SI}).  In a transition matrix, the leading frames are in rows while the following frames are in columns.  There are two behaviors to look for: ``follow-the-leader'' (diagonal elements) where following frames take up the same framing as preceding ones, and ``frame shifting'' (off-diagonal elements) where following frames do not match preceding ones.  

There are two directions of interest: when the instructor's frame precedes the students' frame (figure \ref{fig:IS}), and when the students' frame precedes the instructor's (figure \ref{fig:SI}). Just as in the nested solenoids example, the instructor's participation can come at any point in the problem-solving process.

The instructor-student shift matrix (figure \ref{fig:IS}) tabulates how instructor framing affects student framing.  Follow-the-leader is the dominant mode here: the largest entries are the diagonal ones, and they represent 52\% of all transitions in this matrix.  If frames were uncorrelated with each other, follow-the-leader would occur about 20\% of the time.  

The instructor's marked preference for conceptual physics (CP) is evident in the second row of the matrix, which contains the most entries. Of them, the largest entry is for follow-the-leader behavior: the instructor uses CP and the students continue in CP afterwards.  The other entries in this row constitute frame shifting behavior.
 We interpret the frame shifting behavior to mean that the instructor is just as likely to encourage students into algorithmic frames (AP and AM) as she is to encourage them into ``other'' behavior (e.g. paper shuffling, gossiping, silence).  The former may be because her discussion of conceptual physics matters sets them up to be productive in algorithmic frames, and the latter suggests that perhaps they need more scaffolding -- or a different type of scaffolding -- to be productive problem solvers.

Conversely, we can also investigate how the students' framing affects the instructor's framing.  When the instructor visits each group during problem solving, she first listens to students' ongoing discussion or responds to questions they pose to her.  Because her first interaction with the group is to listen, it's possible that she could mirror their framing (follow-the-leader), or shift it to another frame.  

The students-instructor matrix (figure \ref{fig:SI}) tabulates how student framing leads to instructor framing. In this matrix, follow-the-leader behavior is also common, though no longer dominant: the diagonal terms are large and cover 41\% of the transitions in the matrix. The instructor's preference for conceptual physics (CP) is also apparent in the CP column: no matter what the students start with, if she shifts, it is most likely to be into the CP frame.  This is different than in the instructor-students matrix, where the students rarely shift into this frame if the instructor starts in another frame.  We interpret the asymmetry between these two matrices to be a product of the instructor's ongoing focus on conceptual physics coupled to her more powerful position in the discussion. 

\subsection{Shifts involving three frames}
We can also consider triplets of three frames in a row: the students start in one frame, the instructor responds to that frame and they have a discussion, then the students continue in a third frame.  

If all three frames are the same, the instructor follows the students' lead and they subsequently follow hers. This triplet is the most common. 
% Most often, the instructor provides assistance within their current frame (follow-the-leader, diagonal elements in matrix). 
  Typically, such discussion happens when students start discussing a physics concept and the instructor provides scaffolding to review or strengthen the proposed conceptual argument; this is what happened in the nested solenoids example (section \ref{sec:example}).  Alternately, the students get stuck in computation, the instructor responds with computational help, and the students continue computing.  

Other triplets are possible.  Sometimes the students get stuck in one frame and the instructor responds with new framing, bringing in new information from another frame (often, conceptual physics (CP) as is her wont) to help unstick them.  Usually, they discuss these new ideas with her and can productively return to their original framing. However, on occasion she leaves without completing the discussion and they continue in her framing or lapse into silence. 

\subsection{Shifts involving other codes\label{sec:ocodes}}

In the natural setting of an active physics classroom, there's more than just problem solving. Students and instructors may engage in administrative activities (e.g. discussing or announcing upcoming events, collecting homework) or socializing (e.g. gossiping, discussing sports, developing relationships). These activities are not related to the problem solving process and hence are out of the scope of our framework, even though they are important to the pedagogical purpose of the class and personal relationships of the people within it. Within the scope of this paper, these activities still deserve serious investigation because they may be disruptive to students' problem solving and on-topic framing.  

To investigate how these off-topic activities affect problem solving, we examined the subset of ``other'' codes which pertain to administrative activities or socializing.  Other activities occur frequently during the problem solving process and could involve both students and the instructor as initiators. Overall, they are involved in about a third of all transitions as either the preceding frame, the following frame, or both.

We notice that when the students are engaged in an other activity, the instructor's most common response is to respond in a frame related to problem solving (first row, figure \ref{fig:SI}).  We interpret this as students being off-topic -- possibly because they are stuck -- and the instructor reeling them back in with new ideas. 

Conversely, when the students are in a problem solving frame and the instructor responds in an other activity (first column, figure \ref{fig:SI}), she tends to prefer administrative tasks such as handing back homework.  These responses tend to finish the problem solving episode more often than not, and therefore have no following frame within the episode.   However, when they do not finish the problem solving episode, students are just as likely to follow her off-topic interjection with any of the problem-solving frames. 

\section{Implications}

This paper examined how the instructor's framing affects and is affected by students' framing in an instructional setting with ubiquitous in-class problem solving.  This type of class structure serves our research purposes well but is not typical of upper-division courses in physics.  In related work\cite{Modir2017QuantumFraming}, we used this framework to investigate students' small group problem solving in a quantum mechanics course with a more substantial lecture component.  Students still solved problems in small groups, but problem solving episodes were shorter (usually a few minutes) and less common (only a small handful per week).  Nonetheless, this framework still applied in that class.  We expect that it would also be applicable in other settings where upper-division students work in small groups to solve math-rich physics problems, such as in study groups for homework. 

In this setting, both the students and the instructor followed each other's leads in framing.  This is in marked contrast to the TA frame in \citet{Scherr2009framing}, where a TA joining the group would disrupt students' problem solving.  We suggest that there are two possible causes for this difference.  First, Scherr and Hammer's students are enrolled in an introductory level course and completing tutorials.  Perhaps there is something fundamental about the kind of problems they're solving that causes this difference.  Second, there could be a difference in behavior between their TAs and our instructor.  Our instructor often listened in on students' conversation before joining it, allowing her to pick up on students' framing and follow it; their TAs often spoke immediately upon joining the group.  

As instructors, we can use this framework to pay attention to how students solve problems in our classrooms.  
%When they get stuck, this research suggests that shifting them to or bringing in resources from another frame may be fruitful.   
Taking up this perspective about how students coordinate math and physics ideas conceptually and algorithmically is powerful in its simplicity and broad applicability across many topics at the upper-division. Paying attention to what frames students are in can help us understand how they are coordinating ideas between math and physics, and can help us facilitate their problem solving, whether by supporting their existing framing or by helping to shift them to a new frame.
%Productive problem solving requires shifts between frames as necessary, so noticing student framing and helping students shift frames will support productive problem solving.
There are several concrete implications for instruction using this framework:

%\item When students are stuck, shifting them to another frame may help them become unstuck.  This can be true when they've tied themselves in knots conceptually or when their algorithmic skills are not up to the task at hand.  

As students grind their way through algebraically tedious problems and are moving forward slowly, helping them shift their frame to a conceptual one might help them become productive\cite{Kuo2013Blending}.   Alternately, if students get stuck in problem solving, attending to their framing may help them become unstuck.  This can be true when they've tied themselves in knots conceptually or when their algorithmic skills are not up to the task at hand.  Instructors may choose to shift students to another frame or to help them in their current frame (follow-the-leader), as our instructor did in the solenoids example.

At the end parts of solving a problem, some prescriptive rubrics (e.g. \cite{Heller1992a,Wilcox2015Dirac}) recommend a reflection stage to see if the calculated answer is sensible. This is a golden opportunity for frame shifting: do the calculations agree with conceptual expectations? Alas, it is often treated algorithmically: do the units match? Can I look up this answer on the key? Explicitly thinking about framing can help the instructor tip students\cite{Irving2013framing} into a different kind of reflection. For example, if students are in an algorithmic physics frame (such as to check units), the instructor could ask them to make sense of their solution by thinking about physical reasonableness (conceptual physics) or to compare their functional form to functional forms in similar problems (conceptual mathematics).

Finally, joining students in their existing frame (``follow-the-leader") can be a great way to check in with them and collaboratively nudge them along, supporting their productive problem solving while implicitly (or explicitly) supporting their self-efficacy and agency as learners and problem solvers. These learning goals are not tied to specific physics concepts; nonetheless they are important as we help students try on physics identities. 

Altogether, attending to students' frames as instructors gives us a helpful tool for facilitating groups. 

We've used these four frames here across all of E\&M1 and elsewhere\cite{Modir2017QuantumDifficulties} across quantum mechanics. In contrast, research on student misconceptions or difficulties (e.g. \cite{Pepper1012Observations}) requires that instructors learn myriad different difficulties for each chapter they cover, substantially increasing their professional development overhead and complicating their interactions with students.  Thus, it is a more parsimonious approach for facilitating student learning across topics.  Additionally,  this theoretical framework encourages us to build on students' existing ideas rather than confront and replace them.  In that sense, it is more supportive of students' sense-making during problem solving and may help to increase their self-efficacy in learning physics\cite{Engle2012Expansive,Gupta2013a}.

Just like the instructor in this paper, we often want students to think conceptually about physics problems, integrate ideas between math and physics, and work productively in groups to solve problems.  This theoretical framework helps us as instructors better conceptualize and facilitate student work towards these goals. 
%Throughout the semester, the instructor in this paper often pushed students to the conceptual physics (CP) frame.  From the beginning to the end of the semester, the students became much more likely to use this frame.  This kind of long-term development suggests a deeper and more fundamental kind of learning than mere familiarity with specific physics topics. 

\begin{acknowledgments}

The authors are indebted to insightful colleagues and beta readers who offered criticism on this work. Dylan McKnight identified some of the episodes presented herein and Dina Zohrabi Alaee performed a great deal of our inter-rater reliability testing.

This work was partially supported by the KSU Department of Physics and by NSF grants PHY-1157044 and DUE-1430967. Any opinions, findings, and conclusions or recommendations expressed in this material are those of the authors and do not necessarily reflect the views of the National Science Foundation or other funders.

\end{acknowledgments}

%\bibliography{/Users/le/Dropbox/Research/Bibfiles/library}

%merlin.mbs apsrev4-1.bst 2010-07-25 4.21a (PWD, AO, DPC) hacked
%Control: key (0)
%Control: author (0) dotless jnrlst
%Control: editor formatted (1) identically to author
%Control: production of article title (0) allowed
%Control: page (1) range
%Control: year (0) verbatim
%Control: production of eprint (0) enabled
%

\end{document}